\documentclass[a4paper]{jpconf}
\usepackage{citesort}
\usepackage{graphicx}
\usepackage{iopams}
\usepackage[utf8]{inputenc}
\newcommand{\cO}{{\cal O}}
\newcommand{\cL}{{\cal L}}
\newcommand{\cQ}{{\cal Q}}
\newcommand{\cA}{{\cal A}}

\usepackage{amsmath,bm}
\begin{document}
\title{Isospin-breaking contributions to $\bvarepsilon'/\bvarepsilon$}

\author{V Cirigliano,${}^{1}$ H Gisbert,${}^{2}$ A Pich${}^{3}$ and A Rodr\'iguez-S\'anchez${}^{4,}$\footnote[5]{Speaker}}

\address{${}^{1}$ Theoretical Division, Los Alamos National Laboratory, Los Alamos, NM 87545, USA}

\address{${}^{2}$ Fakult\"at Physik, TU Dortmund, Otto-Hahn-Str. 4, D-44221 Dortmund, Germany}

\address{${}^{3}$  IFIC, Universitat de Val\`encia -- CSIC, Parc Cient{\'i}fic, E-46980 Paterna, Spain}

\address{${}^{4}$  Dep. Astronomy \& Theoretical Physics, Lund U., S\"{o}lvegatan 14A, SE 223-62 Lund, Sweden}

\ead{cirigliano@lanl.gov, hector.gisbert@tu-dortmund.de, antonio.pich@ific.uv.es, antonio.rodriguez@thep.lu.se}

\begin{abstract}
We present an updated analysis of isospin-violating corrections to $\varepsilon'/\varepsilon$ in the framework of chiral perturbation theory, taking advantage of the 
currently improved knowledge on
quark masses and nonperturbative parameters. The role of the different ingredients entering into the analysis is carefully assessed.
Our final result is $\Omega_{\mathrm{eff}}=0.110\,{}^{+0.090}_{-0.088}$ \cite{Cirigliano:2019cpi}.
\end{abstract}

\section{Introduction}
The $K\rightarrow \pi\pi$ process is a splendid laboratory to test our understanding on the interplay of 
the electroweak and strong forces
at different energy regimes \cite{Cirigliano:2011ny}. The initial and final states do not correspond to the fundamental degrees of freedom of the theory, 
{\it i.e.} quarks and gluons. Instead, they are effective Goldstone bosons that emerge from the non-perturbative dynamics of the strong interactions, particularly from its spontaneous chiral symmetry breaking.
However, the process itself involves a change in strangeness which in the Standard Model (SM) can only occur through interactions of weak currents mediated by the $W$ boson, whose mass is several orders of magnitude heavier than $M_K$.
The CKM mechanism introduces a small amount of CP violation that allows for a very small, but different from zero, CP-violating ratio $\varepsilon'/\varepsilon$.

It is convenient to use the following isospin decomposition of the $K\rightarrow \pi\pi$ amplitudes~\cite{Cirigliano:2003gt}:
\begin{eqnarray}  
A(K^0 \to \pi^+ \pi^-) &=&  
\cA_{1/2} + {1 \over \sqrt{2}} \left( \cA_{3/2} + \cA_{5/2} \right) 
\; =\; 
 A_{0}\,  e^{i \chi_0}  + { 1 \over \sqrt{2}}\,    A_{2}\,  e^{i\chi_2 } \, ,
\nonumber\\
A(K^0 \to \pi^0 \pi^0) &=& 
\cA_{1/2} - \sqrt{2} \left( \cA_{3/2} + \cA_{5/2}  \right) 
\; =\;
A_{0}\,  e^{i \chi_0}  - \sqrt{2}\,    A_{2}\,  e^{i\chi_2 }\, ,\label{eq:isodecomp}
\\
A(K^+ \to \pi^+ \pi^0) &=&
{3 \over 2}  \left( \cA_{3/2} - {2 \over 3}\, \cA_{5/2} \right) 
\; =\;
{3 \over 2}\, A_{2}^{+}\,   e^{i\chi_2^{+}},\nonumber
\end{eqnarray}
where $A_0$, $A_2$ and $A_2^+$ are defined to be real in the CP-conserving limit. At first order in CP violation, $\varepsilon'$ can then be written as
\begin{equation}
\varepsilon'\,  =\, 
- \frac{i}{\sqrt{2}} \: e^{i ( \chi_2 - \chi_0 )} \:\omega\;
\left[
\frac{\mathrm{Im} A_{0}}{ \mathrm{Re} A_{0}} \, - \,
\frac{\mathrm{Im} A_{2}}{ \mathrm{Re} A_{2}} \right] 
\, =\, 
- \frac{i}{\sqrt{2}} \: e^{i ( \chi_2 - \chi_0 )} \:\omega\;
\frac{\mathrm{Im} A_{0}}{ \mathrm{Re} A_{0}} \,\bigg( 1\:-\:\frac{1}{\omega}\;\frac{\mathrm{Im} A_{2}}{\mathrm{Im} A_{0}}\bigg) \, ,\nonumber
\end{equation}
where $\omega \equiv\mathrm{Re} A_2 / \mathrm{Re} A_0 \approx 1/22$.

In general, due to the 
small values of the light-quark masses and the electromagnetic coupling, the isospin limit ($m_{d}-m_{u}=\alpha=0$) is a very good approximation. In that limit, $\mathrm{Im}A_{2}$ vanishes. However, a double enhancement destroys this argument for $\varepsilon'/\varepsilon$ \cite{Cirigliano:2003nn,Cirigliano:2003gt,Ecker:2000zr,Cirigliano:2009rr,Donoghue:1986nm,Buras:1987wc,Cheng:1987dk,Lusignoli:1988fz,Wolfe:2000rf}:
\begin{itemize}
\item 
The large enhancement of the $I=0$ amplitude with respect to the $I=2$ one ($1/\omega\approx 22$). 
\item Electroweak penguins diagrams, which are by definition isospin-breaking (IB) contributions, induce
effective $(V - A) \times (V+A)$ four-quark operators 
with enhanced $K\rightarrow \pi\pi$ matrix elements
(their leading-order bosonization does not include derivatives). 
\end{itemize}
The leading role of electroweak penguin contributions to $\mathrm{Im}A_{2}$ is well known. 
Including the rest of IB contributions is crucial  
in order to obtain an accurate prediction for $\varepsilon'/\varepsilon$.
In these proceedings we summarize the results presented in Ref. \cite{Cirigliano:2019cpi}, where 
a complete reanalysis of IB corrections to $\varepsilon'/\varepsilon$ has been performed, updating
the pioneering analysis of \cite{Cirigliano:2003gt,Cirigliano:2003nn}.

\section{From the electroweak scale to the pseudo-Goldstone dynamics}
Non-leptonic kaon decays involve quite different energy regimes.
Taking into account the large value of the strong coupling, logarithm re-summation is a must to go down from the original electroweak scale $\sim M_{W}$ to a mass scale as close as possible to the chiral one, before the perturbative expansion breaks down, {\it i.e.}, $\mu_{SD}\sim 1 \; \mathrm{GeV}$. 
Using the operator product expansion and renormalization group equations, one integrates out the different intermediate degrees of freedom ($m_{t}, M_{W}, m_{b}, m_{c}$), ending up with the effective three-flavour Lagrangian \cite{Buchalla:1995vs}:
\begin{align}
\cL_{\rm eff}^{\Delta S = 1}\:=\:-\frac{G_F}{\sqrt{2}}\:V_{ud}\, V_{us}^*\;\sum_{i=1}^{10} C_i(\mu_{\rm SD})\:Q_i(\mu_{\rm SD})~.\label{eq:shortLagr}
\end{align}
The short-distance information is encoded in the perturbative Wilson Coefficients $C_{i}(\mu_{SD})$,
which are known to next-to-leading order (NLO) \cite{Buras:1991jm,Buras:1992tc,Buras:1992zv,Ciuchini:1993vr}. The complete calculation of next-to-next-to-leading-order (NNLO) QCD corrections is expected to be finished soon \cite{Cerda-Sevilla:2016yzo,Buras:1999st,Gorbahn:2004my}. 

The long-distance dynamics is contained in the hadronic matrix elements $\langle \pi\pi | Q_i(\mu_{\rm SD}) | K\rangle$. In order to analytically evaluate them, one has to rewrite the relevant operators in the effective low-energy theory of the strong interactions, 
imposing that they must transform in the same way under $SU(3)_{L}\times SU(3)_{R}$. The introduction of those operators adds to the usual strong \cite{Gasser:1984gg,Fearing:1994ga,Bijnens:1999sh} and electromagnetic chiral-perturbation-theory ($\chi\mathrm{PT}$) Lagrangians \cite{Ecker:1988te,Urech:1994hd},
\begin{align}
\cL_{\rm strong}  \; &=\; \frac{F^2}{4} \,\langle D_\mu U D^\mu U^\dagger + 
\chi U^\dagger + \chi^\dagger  U \rangle
 +  \sum_{i=1}^{10}\; L_i\, O^{p^4}_i +  F^{-2}\sum_{i=1}^{90}\; X_i\, O^{p^6}_i
 + \cO(p^8)~, 
 \label{eq:Lstrong}\\
\cL_{\rm elm} \; &=\;  e^2\, Z \, F^4  \,\langle \cQ  U^\dagger \cQ U\rangle 
+ e^2 \,F^2\, \sum^{14}_{i=1}\; K _i \, O^{e^2 p^2}_i + \cO(e^2p^4)~, 
\label{eq:Lelm}
\end{align}
the weak nonleptonic \cite{Cronin:1967jq,Kambor:1989tz,Ecker:1992de,Bijnens:1998mb} and electroweak ones \cite{Bijnens:1983ye,Grinstein:1985ut,Ecker:2000zr}:
\begin{align} \nonumber
\cL^{\Delta S=1}
&=  G_8 \, F^4  \,\langle\lambda D^\mu U^\dagger
 D_\mu U \rangle  + G_8\, F^2\sum^{22}_{i=1}   N_i \, O^8_i \nonumber\\
& +  G_{27}\, F^4 \left( L_{\mu 23} L^\mu_{11} + 
{2\over 3} L_{\mu 21} L^\mu_{13}\right)  +
  G_{27}\, F^2\sum^{28}_{i=1} D_i \, O^{27}_i + \cO(G_{F} p^6)\, , 
\label{eq:Lweak}\\
\cL^{\Delta S=1}_{\rm EW} \; &=\; e^2 \,G_8 \, g_{\rm ewk}\, F^6 \,\langle\lambda U^\dagger \cQ 
U\rangle 
 +   e^2   \, G_8\, F^4\,\sum^{14}_{i=1} Z_i\, O^{EW}_i + \cO(G_F e^2 p^4)\, .
\label{eq:Lelweak} 
\end{align}
The matrix $U(x)\equiv\exp{\{i \lambda^a \phi^a(x)/F\}}$ parametrizes the pseudoscalar Goldstone fields. The definitions of the rest of terms can be found in Ref. \cite{Cirigliano:2019cpi}. 

When one bosonizes the operators by simply using their symmetry transformations, there is an ambiguous relative normalization. The normalization of the quark currents becomes fixed at leading chiral order by gauge invariance and/or by fitting the meson masses. This is no longer the case for the four-quark operators and, as a consequence, the values of the different low-energy constants (LECs) $G_{8}, G_{27}, G_{8}g_{ewk}, \{N_{i}\}, \{D_{i}\}, \{Z_{i}\}$, which can be regarded as the Wilson Coefficients of $\chi$PT, are not known. They encode the information on the dynamics above the chiral scale $\Lambda_{\chi}$ and their values should be extracted from the short-distance Lagrangian  (\ref{eq:shortLagr}). However, we do not know how to analytically perform such a non-perturbative matching.

In order to estimate the numerical values of the LECs we can make use of the fact that in
the limit of a large number of QCD colours the T-product of two colour-singlet quark currents factorizes. Since we know how the quark currents bosonize, normalization included, the large-$N_{c}$ matching of Eqs. (\ref{eq:Lweak}) and (\ref{eq:Lelweak}) with Eq. (\ref{eq:shortLagr}) becomes then possible, which determines the weak LECs as functions of the better-known strong and electromagnetic ones in Eqs. (\ref{eq:Lstrong}) and (\ref{eq:Lelm}).

Compact analytical results for this large-$N_{c}$ matching
can be found in Refs. \cite{Cirigliano:2019cpi,Cirigliano:2003gt}. 
It is important to remark that the large-$N_C$ limit is only used to fix the values of the LECs. Our calculation of the $K\to\pi\pi$ amplitudes incorporates all NLO $\chi$PT corrections, which include very important logarithmic contributions of $\mathcal{O}(1/N_{c})$ \cite{Gisbert:2017vvj,Pallante:2001he,Pallante:2000hk,Pallante:1999qf}. 

\section{Reassessment of the numerical inputs}
One of the main motivations for this reanalysis is the large number of improvements in the needed inputs, since the original study of Ref.~\cite{Cirigliano:2003gt}. The more relevant physical parameters are:
\begin{itemize}
\item \textbf{Quark masses and Wilson Coefficients:} $m_{s}(\mu_{\rm SD}=1\, \mathrm{GeV})=125.6 \pm 0.9_{m_{s}} \pm 1.9_{\alpha_{s}} \, \mathrm{MeV}$ and $m_{d}(\mu_{\rm SD}=1\, \mathrm{GeV})=6.27 \pm 0.12_{m_{d}} \pm 0.09_{\alpha_{s}} \, \mathrm{MeV}$~\cite{Aoki:2019cca}.
The inputs for obtaining the Wilson coefficients are taken from \cite{Aoki:2019cca,Aad:2019mkw}.
We use two different definitions of $\gamma_5$ (naive dimensional regularisation and 't Hooft-Veltman \cite{tHooft:1972tcz}) and average the two results.

\item \textbf{$\mathbf{O(p^{4})}$ strong LECs:} 
The most recent determinations from continuum fits \cite{Bijnens:2014lea} and lattice simulations \cite{Aoki:2019cca} are in good agreement with the large-$N_{c}$ resonance-saturation estimates \cite{Ecker:1988te,Pich:2002xy}. We adopt the values (in $10^{-3}$ units):
\begin{equation}\label{eq:Li-inputs}
\begin{array}{ccc}
L_5^r(M_\rho) = 1.20\pm 0.10\; ,  \qquad  
&L_7 = -0.32\pm 0.10 \; ,& \qquad
L_8^r(M_\rho) = 0.53\pm 0.11 \; . 
\end{array}
\end{equation}

\item \textbf{$\mathbf{O(p^{6})}$ strong LECs:} A complete analysis of resonance contributions to the $\mathcal{O}(p^{6})$ LECs was made in Ref. \cite{Cirigliano:2006hb} by matching the $\chi$PT and R$\chi$T Lagrangians at leading order (LO) in $1/N_C$, in the single-resonance approximation. The $X_{i}$ LECs relevant for $K\to\pi\pi$
only receive contributions from scalar and pseudoscalar resonance-exchange.

\item \textbf{Electromagnetic LECs:} Our numerical inputs for the couplings $K_i$ are based on $\chi$PT and R$\chi$T analyses of different two-, three- and four-point Green functions \cite{Moussallam:1997xx,Ananthanarayan:2004qk,Cirigliano:2003gt,Bijnens:1996kk,Albaladejo:2017hhj}.
\item \textbf{Phenomenological fit of the CP-even amplitudes:} The CP-even parts of $A_I$, together with the phase-shift difference $\chi_{0}-\chi_{2}$, can be directly extracted from the measured partial widths $\Gamma_{+-,00,+0}$ \cite{PhysRevD.98.030001}. Including NLO $\chi$PT corrections and IB effects, this allows for an empirical determination of $\mathrm{Re}(G_{8,27})$. The large-$N_C$ matching is only needed
for the CP-odd parts of the weak LECs, which are dominated by $Q_{6}$ and $Q_{8}$. Since these two operators have non-vanishing anomalous dimensions at $N_C\to\infty$, their matrix elements are well approximated in this limit.
\end{itemize}
\section{$\mathbf{K\boldsymbol{\to}\bpi\bpi}$ amplitudes up to NLO}
The LO amplitudes in the CP conserving limit,
\begin{eqnarray}
\cA_{1/2} & = &  - \sqrt{2}\, G_8 F\, \Big[  \left( M_{K}^2 - M_{\pi}^2 \right)
 \Big] - {\sqrt{2} \over 9}\, G_{27} F \left( M_{K}^2 - M_{\pi}^2 \right)  ,\nonumber
\\
\cA_{3/2} & = & -
{10 \over 9}\,  G_{27} F \left( M_{K}^2 - M_{\pi}^2 \right),\label{eq:LOamp}\\
\cA_{5/2} & = & 0\, ,\nonumber
\end{eqnarray}
are altered by IB corrections in three different ways:
\begin{enumerate}
\item \textbf{$\mathbf{\bpi^{0}}$-$\bfeta$ mixing.} Taking into account that $m_{u}\neq m_{d}$, the pseudoscalar fields $\phi^{3}$ and $\phi^{8}$ in Eq.~(\ref{eq:Lstrong}) are not mass-eigenstates anymore.
The field re-definition leads to extra terms.
\item \textbf{Mass corrections.} They arise both from the $m_{d}-m_{u}$ mass difference and from electromagnetism through the LEC $Z$ of Eq. (\ref{eq:Lelm}).
\item \textbf{Electroweak Lagrangian.} They arise from inserting one vertex from Eq. (\ref{eq:Lelweak}) instead of one from Eq. (\ref{eq:Lweak}).
\end{enumerate}

At NLO there are different kinds of IB contributions. The light-quark mass difference and the electromagnetic interaction induce corrections through the meson masses (both when putting them on-shell and in the propagators) and $\pi^{0}$-$\eta$ mixing at NLO. 
The presence of electromagnetism introduces also tree-level diagrams with at least one electroweak vertex and a NLO insertion, loop corrections with one $G_{8}g_{ewk}$ vertex, photon loops and radiative counterparts. Taking all of them into account, the chiral amplitudes $\cA_n$ ($n=1/2,\:3/2,\:5/2$) at NLO, including IB, can be expressed as:
\begin{align}
\cA_n &= - G_{27} \, F_\pi \, \Big( M_K^2 - M_\pi^2 \Big) \,  \cA_{n}^{(27)} - 
G_{8} \, F_\pi \,
\Big( M_K^2 - M_\pi^2 \Big) 
\Big[ \cA_{n}^{(8)} +  \varepsilon^{(2)} \, \cA_{n}^{(\varepsilon)} \Big] \nonumber\\
& +   e^2 \,G_{8} \, F_\pi^3 \,  \Big[ \cA_{n}^{(\gamma)} + Z \,  \cA_{n}^{(Z)} + 
 g_{\rm ewk} \,  \cA_{n}^{(g)} \Big] ~,\label{eq:generalamplitude}
\end{align}
where, in general, every amplitude $\cA_{n}^{(X)}$ can be decomposed as the sum of a tree-level, a loop and a counter-term part. Our numerical results are displayed in Tables 4, 5 and 6 of Ref. \cite{Cirigliano:2019cpi}.

\section{Isospin-breaking terms in $\bvarepsilon'/\bvarepsilon$}
At first order in isospin breaking~\cite{Cirigliano:2003gt,Cirigliano:2003nn},
\begin{equation} 
\varepsilon' = - \frac{i}{\sqrt{2}} \, e^{i ( \chi_2 - \chi_0 )} \, 
\omega_+  \,   \left[ 
\frac{\mathrm{Im} A_{0}^{(0)}}{ \mathrm{Re} A_{0}^{(0)}} \, 
(1 - \Omega_{\mathrm{eff}})
- \frac{\mathrm{Im} A_{2}^{\mathrm{emp}}}{ \mathrm{Re}
  A_{2}^{(0)}} \right] , 
\label{eq:cpiso}
\end{equation}
where the $(0)$ superscript indicates the isospin limit and the IB correction contains three effects:
\begin{equation}
\Omega_{\mathrm{eff}}\,\equiv\, \Omega_{\mathrm{IB}}-\Delta_{0}-f_{5/2}\, .
\end{equation}
$f_{5/2}$ arises from rewriting $\omega$ in terms of the experimentally known $\omega_{+}\equiv \mathrm{Re}A_{2}^{+}/\mathrm{Re}\mathrm{A_{0}}$.
$\Delta_{0}$ includes all the IB corrections to $A_{0}$ and
$\Omega_{\mathrm{IB}}$ contains the $\mathrm{Im} A_{2}$ pieces that do not come from electromagnetic penguins.

The main results of our analysis are displayed in Table \ref{tab:IB-results}, where the estimated values for the IB parameters are shown at different levels of approximation. The hierarchy of the different corrections agrees with the  expectations. The dominant IB contribution comes from $\Omega_{\text{IB}}$, which contains those corrections enhanced by $1/\omega$. Both the light-quark mass difference (corresponding in the table to the $\alpha=0$ value) and those electromagnetic corrections that do not come from electromagnetic penguins induce corrections of approximately the same size. The IB corrections to $A_{0}$ are dominated by the electromagnetic contributions, which contain the chirally-enhanced electroweak penguins.\footnote{Those otherwise isospin-conserving analyses that also incorporate the electroweak penguin contributions to $A_{0}$ should subtract them from $\Delta_{0}$. 
The corresponding IB parameter is
$\hat{\Omega}_{\text{eff}}\:\equiv\:\Omega_{\text{IB}}\:-\:\Delta_0|_{\alpha=0}\:-\:f_{5/2}=(17.0\:{}^{+9.1}_{-9.0})\cdot 10^{-2}$.} The relatively large value of $f_{5/2}$ may be understood once again from the numerically small $\mathrm{Re}A_{2}^{(0)}$, provided their IB terms do not present the same overall suppression. The sum of the three contributions happens to be destructive, leading to our final value:
\begin{table}[t]
\centering
\begin{tabular}{|c|cc|cc|}
\hline
& \multicolumn{2}{|c|}{$\alpha =0$} & \multicolumn{2}{|c|}{$\alpha\not= 0$}
\\
& LO & NLO  & LO & NLO
\\ \hline
$\Omega_{\text{IB}}$ & $13.7$  & $15.9\pm 8.2$ & $19.5\pm 3.9$ & $24.7\pm 7.8$
\\
$\Delta_{0}$ & $-0.002$ & $-0.49\pm 0.13$ & $5.6 \pm 0.9$ & $5.6\pm 0.9$
\\
$f_{5/2}$ & 0 & 0 & 0 & $8.2\, ^{\,+\,2.3}_{\,-\,2.5}$
\\ \hline
$\Omega_{\text{eff}}$ & $13.7$ & $16.4\pm 8.3$ & $13.9\pm 3.7$ & $11.0\, {^{\,+\,9.0}_{\,-\,8.8}}$
\\ \hline
\end{tabular}
\caption{Isospin-violating corrections for $\epsilon'/\epsilon$ in units of $10^{-2}$. }
\label{tab:IB-results}
\end{table}
\begin{equation}
\Omega_{\text{eff}}\, =\, 0.110\,{}^{+0.090}_{-0.088} \, .
\end{equation}
Uncertainties have been estimated conservatively, taking into account the errors of the different inputs (see Ref. \cite{Cirigliano:2019cpi} for details). The final one is dominated both by the uncertainty in the $\cO(p^{4})$ coupling $L_{7}$ and by the systematic uncertainty arising from the ambiguity when fixing the chiral scale $\nu_{\chi}$, which parametrizes our ignorance on $1/N_{c}$ corrections. 

Taking into account the correlations, the associated SM prediction for $\varepsilon'/\varepsilon$ \cite{Cirigliano:2019cpi,Gisbert:2017vvj,Pallante:2001he} becomes
\begin{align}\label{eq:epsp_prediction}
\text{Re}\left(\epsilon'/\epsilon\right)=\:\left(14\,\pm\,5\right)\cdot 10^{-4} \, ,
\end{align}
in good agreement with the experimental world average \cite{Batley:2002gn,Lai:2001ki,Fanti:1999nm,Barr:1993rx,Burkhardt:1988yh,Abouzaid:2010ny,AlaviHarati:2002ye,AlaviHarati:1999xp,Gibbons:1993zq}.

\section*{Acknowledgements}
We warmly acknowledge early collaboration with Gerhard Ecker and Helmut Neufeld. This work has been supported in part by the Spanish Government and ERDF funds from
the EU Commission [grant FPA2017-84445-P], the Generalitat Valenciana [grant Prometeo/2017/053], the Swedish Research Council [grants 2015-04089  and  2016-05996]  and  the  European  Research Council (ERC) under the EU Horizon 2020 research and innovation programme (grant 668679). The work of H.G. is supported by 
the Bundesministerium f\"ur Bildung und Forschung (BMBF).
V.C. acknowledges support by the US DOE Office of Nuclear Physics.

\section*{References}
\bibliography{iopart-num}

\providecommand{\newblock}{}
\begin{thebibliography}{10}
\expandafter\ifx\csname url\endcsname\relax
  \def\url#1{{\tt #1}}\fi
\expandafter\ifx\csname urlprefix\endcsname\relax\def\urlprefix{URL }\fi
\providecommand{\eprint}[2][]{\url{#2}}

\bibitem{Cirigliano:2019cpi}
Cirigliano V, Gisbert H, Pich A and Rodr\'{\i}guez-S\'{a}nchez A 2019
  (\textit{Preprint} \eprint{1911.01359})

\bibitem{Cirigliano:2011ny}
Cirigliano V, Ecker G, Neufeld H, Pich A and Portoles J 2012 {\em Rev. Mod.
  Phys.\/} {\bf 84} 399 (\textit{Preprint} \eprint{1107.6001})

\bibitem{Cirigliano:2003gt}
Cirigliano V, Ecker G, Neufeld H and Pich A 2004 {\em Eur. Phys. J.\/} {\bf
  C33} 369--396 (\textit{Preprint} \eprint{hep-ph/0310351})

\bibitem{Cirigliano:2003nn}
Cirigliano V, Pich A, Ecker G and Neufeld H 2003 {\em Phys. Rev. Lett.\/} {\bf
  91} 162001 (\textit{Preprint} \eprint{hep-ph/0307030})

\bibitem{Ecker:2000zr}
Ecker G, Isidori G, Muller G, Neufeld H and Pich A 2000 {\em Nucl. Phys.\/}
  {\bf B591} 419--434 (\textit{Preprint} \eprint{hep-ph/0006172})

\bibitem{Cirigliano:2009rr}
Cirigliano V, Ecker G and Pich A 2009 {\em Phys. Lett.\/} {\bf B679} 445--448
  (\textit{Preprint} \eprint{0907.1451})

\bibitem{Donoghue:1986nm}
Donoghue J~F, Golowich E, Holstein B~R and Trampetic J 1986 {\em Phys. Lett.\/}
  {\bf B179} 361 [Erratum: Phys. Lett.B188,511(1987)]

\bibitem{Buras:1987wc}
Buras A~J and Gerard J~M 1987 {\em Phys. Lett.\/} {\bf B192} 156--162

\bibitem{Cheng:1987dk}
Cheng H~Y 1988 {\em Phys. Lett.\/} {\bf B201} 155--159

\bibitem{Lusignoli:1988fz}
Lusignoli M 1989 {\em Nucl. Phys.\/} {\bf B325} 33--61

\bibitem{Wolfe:2000rf}
Wolfe C~E and Maltman K 2001 {\em Phys. Rev.\/} {\bf D63} 014008
  (\textit{Preprint} \eprint{hep-ph/0007319})

\bibitem{Buchalla:1995vs}
Buchalla G, Buras A~J and Lautenbacher M~E 1996 {\em Rev. Mod. Phys.\/} {\bf
  68} 1125--1144 (\textit{Preprint} \eprint{hep-ph/9512380})

\bibitem{Buras:1991jm}
Buras A~J, Jamin M, Lautenbacher M~E and Weisz P~H 1992 {\em Nucl. Phys.\/}
  {\bf B370} 69--104 [Addendum: Nucl. Phys.B375,501(1992)]

\bibitem{Buras:1992tc}
Buras A~J, Jamin M, Lautenbacher M~E and Weisz P~H 1993 {\em Nucl. Phys.\/}
  {\bf B400} 37--74 (\textit{Preprint} \eprint{hep-ph/9211304})

\bibitem{Buras:1992zv}
Buras A~J, Jamin M and Lautenbacher M~E 1993 {\em Nucl. Phys.\/} {\bf B400}
  75--102 (\textit{Preprint} \eprint{hep-ph/9211321})

\bibitem{Ciuchini:1993vr}
Ciuchini M, Franco E, Martinelli G and Reina L 1994 {\em Nucl. Phys.\/} {\bf
  B415} 403--462 (\textit{Preprint} \eprint{hep-ph/9304257})

\bibitem{Cerda-Sevilla:2016yzo}
Cerd\`a-Sevilla M, Gorbahn M, Jäger S and Kokulu A 2017 {\em J. Phys. Conf.
  Ser.\/} {\bf 800} 012008 (\textit{Preprint} \eprint{1611.08276})

\bibitem{Buras:1999st}
Buras A~J, Gambino P and Haisch U~A 2000 {\em Nucl. Phys.\/} {\bf B570}
  117--154 (\textit{Preprint} \eprint{hep-ph/9911250})

\bibitem{Gorbahn:2004my}
Gorbahn M and Haisch U 2005 {\em Nucl. Phys.\/} {\bf B713} 291--332
  (\textit{Preprint} \eprint{hep-ph/0411071})

\bibitem{Gasser:1984gg}
Gasser J and Leutwyler H 1985 {\em Nucl. Phys.\/} {\bf B250} 465--516

\bibitem{Fearing:1994ga}
Fearing H~W and Scherer S 1996 {\em Phys. Rev.\/} {\bf D53} 315--348
  (\textit{Preprint} \eprint{hep-ph/9408346})

\bibitem{Bijnens:1999sh}
Bijnens J, Colangelo G and Ecker G 1999 {\em JHEP\/} {\bf 02} 020
  (\textit{Preprint} \eprint{hep-ph/9902437})

\bibitem{Ecker:1988te}
Ecker G, Gasser J, Pich A and de~Rafael E 1989 {\em Nucl. Phys.\/} {\bf B321}
  311--342

\bibitem{Urech:1994hd}
Urech R 1995 {\em Nucl. Phys.\/} {\bf B433} 234--254 (\textit{Preprint}
  \eprint{hep-ph/9405341})

\bibitem{Cronin:1967jq}
Cronin J~A 1967 {\em Phys. Rev.\/} {\bf 161} 1483--1494

\bibitem{Kambor:1989tz}
Kambor J, Missimer J~H and Wyler D 1990 {\em Nucl. Phys.\/} {\bf B346} 17--64

\bibitem{Ecker:1992de}
Ecker G, Kambor J and Wyler D 1993 {\em Nucl. Phys.\/} {\bf B394} 101--138

\bibitem{Bijnens:1998mb}
Bijnens J, Pallante E and Prades J 1998 {\em Nucl. Phys.\/} {\bf B521} 305--333
  (\textit{Preprint} \eprint{hep-ph/9801326})

\bibitem{Bijnens:1983ye}
Bijnens J and Wise M~B 1984 {\em Phys. Lett.\/} {\bf 137B} 245--250

\bibitem{Grinstein:1985ut}
Grinstein B, Rey S~J and Wise M~B 1986 {\em Phys. Rev.\/} {\bf D33} 1495

\bibitem{Gisbert:2017vvj}
Gisbert H and Pich A 2018 {\em Rept. Prog. Phys.\/} {\bf 81} 076201
  (\textit{Preprint} \eprint{1712.06147})

\bibitem{Pallante:2001he}
Pallante E, Pich A and Scimemi I 2001 {\em Nucl. Phys.\/} {\bf B617} 441--474
  (\textit{Preprint} \eprint{hep-ph/0105011})

\bibitem{Pallante:2000hk}
Pallante E and Pich A 2001 {\em Nucl. Phys.\/} {\bf B592} 294--320
  (\textit{Preprint} \eprint{hep-ph/0007208})

\bibitem{Pallante:1999qf}
Pallante E and Pich A 2000 {\em Phys. Rev. Lett.\/} {\bf 84} 2568--2571
  (\textit{Preprint} \eprint{hep-ph/9911233})

\bibitem{Aoki:2019cca}
Aoki S {\em et~al.\/} (Flavour Lattice Averaging Group) 2019
  (\textit{Preprint} \eprint{1902.08191})

\bibitem{Aad:2019mkw}
Aad G {\em et~al.\/} (ATLAS) 2019  (\textit{Preprint} \eprint{1905.02302})

\bibitem{tHooft:1972tcz}
't~Hooft G and Veltman M~J~G 1972 {\em Nucl. Phys.\/} {\bf B44} 189--213

\bibitem{Bijnens:2014lea}
Bijnens J and Ecker G 2014 {\em Ann. Rev. Nucl. Part. Sci.\/} {\bf 64} 149--174
  (\textit{Preprint} \eprint{1405.6488})

\bibitem{Pich:2002xy}
Pich A 2002 {\em {Phenomenology of large $N_C$ QCD. Proceedings, Tempe, USA,
  January 9-11, 2002}\/} pp 239--258 (\textit{Preprint}
  \eprint{hep-ph/0205030})

\bibitem{Cirigliano:2006hb}
Cirigliano V, Ecker G, Eidemuller M, Kaiser R, Pich A and Portoles J 2006 {\em
  Nucl. Phys.\/} {\bf B753} 139--177 (\textit{Preprint}
  \eprint{hep-ph/0603205})

\bibitem{Moussallam:1997xx}
Moussallam B 1997 {\em Nucl. Phys.\/} {\bf B504} 381--414 (\textit{Preprint}
  \eprint{hep-ph/9701400})

\bibitem{Ananthanarayan:2004qk}
Ananthanarayan B and Moussallam B 2004 {\em JHEP\/} {\bf 06} 047
  (\textit{Preprint} \eprint{hep-ph/0405206})

\bibitem{Bijnens:1996kk}
Bijnens J and Prades J 1997 {\em Nucl. Phys.\/} {\bf B490} 239--271
  (\textit{Preprint} \eprint{hep-ph/9610360})

\bibitem{Albaladejo:2017hhj}
Albaladejo M and Moussallam B 2017 {\em Eur. Phys. J.\/} {\bf C77} 508
  (\textit{Preprint} \eprint{1702.04931})

\bibitem{PhysRevD.98.030001}
Tanabashi M {\em et~al.\/} (Particle Data Group) 2018 {\em Phys. Rev.\/} {\bf
  D98} 030001

\bibitem{Batley:2002gn}
Batley J~R {\em et~al.\/} (NA48) 2002 {\em Phys. Lett.\/} {\bf B544} 97--112
  (\textit{Preprint} \eprint{hep-ex/0208009})

\bibitem{Lai:2001ki}
Lai A {\em et~al.\/} (NA48) 2001 {\em Eur. Phys. J.\/} {\bf C22} 231--254
  (\textit{Preprint} \eprint{hep-ex/0110019})

\bibitem{Fanti:1999nm}
Fanti V {\em et~al.\/} (NA48) 1999 {\em Phys. Lett.\/} {\bf B465} 335--348
  (\textit{Preprint} \eprint{hep-ex/9909022})

\bibitem{Barr:1993rx}
Barr G~D {\em et~al.\/} (NA31) 1993 {\em Phys. Lett.\/} {\bf B317} 233--242

\bibitem{Burkhardt:1988yh}
Burkhardt H {\em et~al.\/} (NA31) 1988 {\em Phys. Lett.\/} {\bf B206} 169--176

\bibitem{Abouzaid:2010ny}
Abouzaid E {\em et~al.\/} (KTeV) 2011 {\em Phys. Rev.\/} {\bf D83} 092001
  (\textit{Preprint} \eprint{1011.0127})

\bibitem{AlaviHarati:2002ye}
Alavi-Harati A {\em et~al.\/} (KTeV) 2003 {\em Phys. Rev.\/} {\bf D67} 012005
  [Erratum: Phys. Rev.D70,079904(2004)] (\textit{Preprint}
  \eprint{hep-ex/0208007})

\bibitem{AlaviHarati:1999xp}
Alavi-Harati A {\em et~al.\/} (KTeV) 1999 {\em Phys. Rev. Lett.\/} {\bf 83}
  22--27 (\textit{Preprint} \eprint{hep-ex/9905060})

\bibitem{Gibbons:1993zq}
Gibbons L~K {\em et~al.\/} 1993 {\em Phys. Rev. Lett.\/} {\bf 70} 1203--1206

\end{thebibliography}

\end{document}